\documentclass[%
reprint,
superscriptaddress,
 amsmath,amssymb,
 aps,
pra,
longbibliography,
floatfix,
]{revtex4-1}
\usepackage{graphicx}
\usepackage{color}
\usepackage{hyperref}
\usepackage{float}
\usepackage{array, multirow}
\usepackage{amsmath}
\usepackage{newtxtext}              \usepackage{booktabs}
\usepackage{xcolor} 
\usepackage{soul}
\usepackage{url}
\usepackage[margin=0.8in]{geometry} 

\hypersetup{colorlinks, 
    linkcolor={blue!75!black!80!yellow},
    citecolor={blue!75!black!80!yellow}, 
    urlcolor={blue!75!black!80!yellow}
    }
\begin{document}

\title{Capturing Non-Markovian Dynamics on Near-Term Quantum Computers}

\author{Kade Head-Marsden}
\email{kheadmarsden@seas.harvard.edu}
\affiliation{John A. Paulson School of Engineering and Applied Sciences, Harvard
University, Cambridge, MA 02138, USA}
\author{Stefan Krastanov}
\affiliation{John A. Paulson School of Engineering and Applied Sciences, Harvard
University, Cambridge, MA 02138, USA}
\affiliation{Department of Electrical Engineering and Computer Science, Massachusetts Institute of Technology, Cambridge, MA 02139, USA}
\author{David A. Mazziotti}
\affiliation{Department of Chemistry and The James Franck Institute, The University of Chicago, Chicago, IL 60637 USA}
\author{Prineha Narang}
\email{prineha@seas.harvard.edu}
\affiliation{John A. Paulson School of Engineering and Applied Sciences, Harvard
University, Cambridge, MA 02138, USA}

\date{\today}

\begin{abstract}
\noindent With the rapid progress in quantum hardware, there has been an increased interest in new quantum algorithms to describe complex many-body systems searching for the still-elusive goal of `useful quantum advantage'. Surprisingly, quantum algorithms for the treatment of open quantum systems (OQSs) have remained under-explored, in part due to the inherent challenges of mapping non-unitary evolution into the framework of unitary gates. Evolving an open system unitarily necessitates dilation into a new effective system to incorporate critical environmental degrees of freedom. In this context, we present and validate a new quantum algorithm to treat non-Markovian dynamics in OQSs built on the Ensemble of Lindblad's Trajectories approach, invoking the Sz.-Nagy dilation theorem. Here we demonstrate our algorithm on the Jaynes-Cummings model in the strong coupling and detuned regimes, relevant in quantum optics and driven quantum systems studies. This algorithm, a key step towards generalized modeling of non-Markovian dynamics on a noisy-quantum device, captures a broad class of dynamics and opens up a new direction in OQS problems. 
\end{abstract}

\maketitle
\section{Introduction}
\noindent Open quantum systems (OQSs), quantum systems that are coupled to their environment, are ubiquitous in the physical sciences~\cite{Gardiner2004, Maniscalco2006, Barreiro2011, Mazziotti2012, Skochdopole2014, Chakraborty2015, Chakraborty2017, Fetherolf2017, Flick2018, Flick2018b} and many classical techniques exist to describe the dynamics of an OQS beyond the Markov approximation~\cite{Breuer2002, Shabani2005, Ishizaki2005, Chen2009, Chruscinski2010, Barchielli2012, Budini2014, Montoya-Castillo2015, Vacchini2016, Breuer2016, deVega2017,Campbell2018, HeadMarsden2019PRA, HeadMarsden2019JCP, Reimer2019, Yoshioka2019, Nagy2019, Hartmann2019, Vicentini2019}. While the Lindblad formalism gives an efficient and accurate depiction of the dynamics in the weak coupling regime~\cite{Lindblad1976, Gorini1976}, the approach does not extend to systems that are strongly coupled to their environments~~\cite{Breuer2002,deVega2017}. Strong coupling can lead to non-Markovian effects such as recurrences of quantum properties, which are both important for a fundamental understanding of system dynamics and show promise for aiding in system control~\cite{Wu2007, Rebentrost2009, Pachon2013, Laine2014, Reich2015, Poggi2017, Ho2019, Mirkin2019a, Mirkin2019b}. With the advent of quantum devices and the corresponding search for useful quantum advantage there has been an increased interest in algorithm development for physics and chemistry problems. Surprisingly, algorithm development for the treatment of open quantum systems has been limited to a few theoretical and experimental studies~\cite{Barreiro2011, Wang2011, Sweke2015, Wei2016, Hu2020, Liu2020, Garcia-Perez2020}. While there have been impressive recent strides in the field,~\cite{Hu2020, Garcia-Perez2020} the lag in the development of this field is in part due to the challenge of the non-unitary evolution of OQSs being cast into the framework of unitary quantum gates. To evolve an OQS unitarily, dilation methods must be used to incorporate the important environmental degrees of freedom into a new effective system. Early work faced computational scaling challenges with this dilation~\cite{Sweke2015}; however, recent advances~\cite{Langer1972, Paulsen2002, Levy2014} have allowed accurate simulation of Lindbladian dynamics on a noisy-quantum device~\cite{Hu2020}.

In this \textcolor{black}{\emph{Article}}, we present a novel quantum algorithm to treat non-Markovian dynamics in open quantum systems by extending the Ensemble of Lindblad's Trajectories method onto a quantum computer (ELT-QC) by invoking the Sz.-Nagy dilation theorem~\cite{Langer1972, Paulsen2002, Levy2014, Hu2020}. While previous work has looked at example systems for Markovian and non-Markovian dynamics on quantum computers, here we aim to provide the foundation for a universal and  generalizable theory that allows for the investigation into open quantum system properties. We start with introducing the theory behind the ELT-QC method, followed by benchmarking the algorithm, and demonstrating the impact of this approach on problems in quantum optics.

\section{Theory}
\label{sec:theory}
\subsection{Ensemble of Lindbladian Trajectories}
\begin{figure}[h!]
 \includegraphics[width=0.5\textwidth]{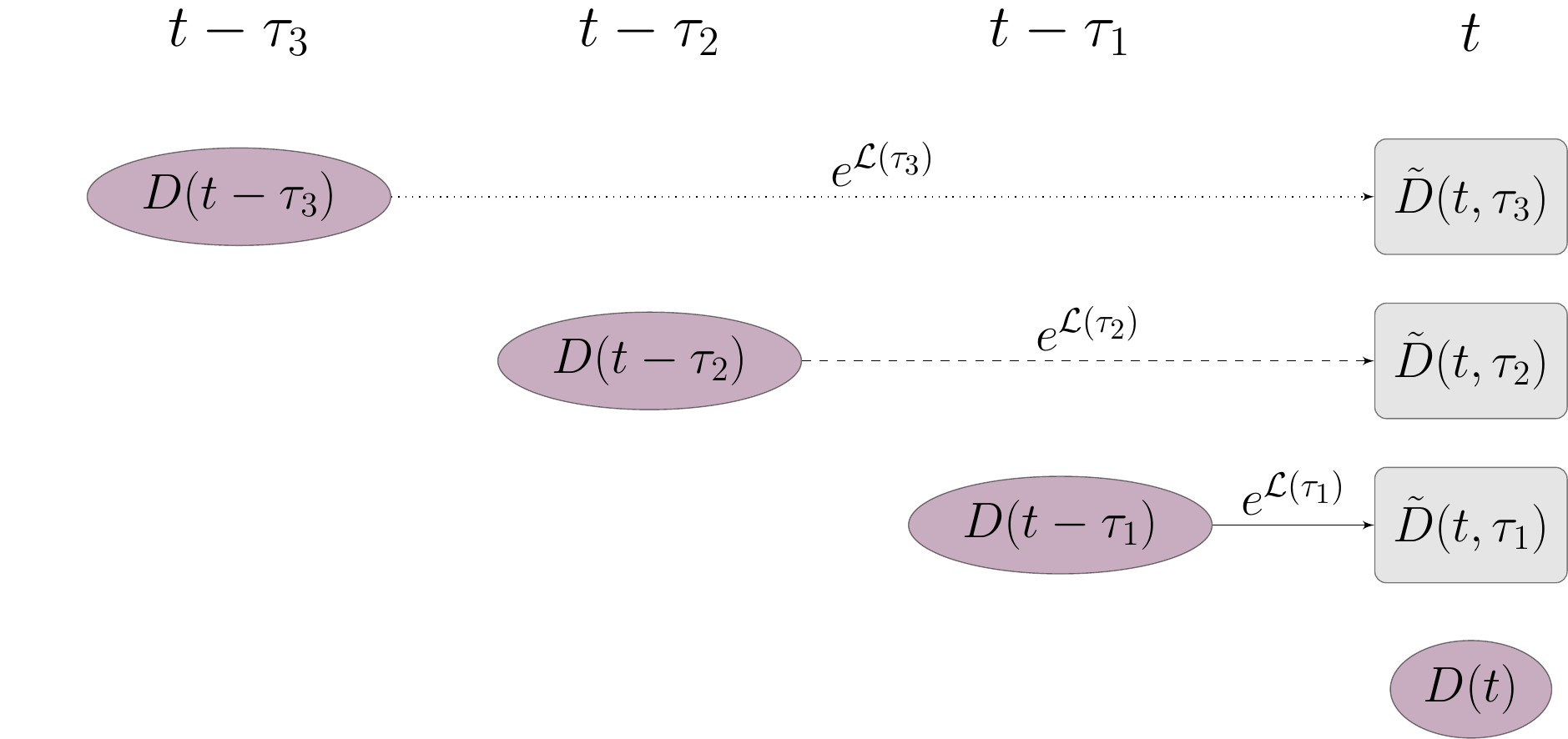}\\
 \caption{An ensemble of Lindbladian trajectories whose weighted ensemble produces the true density matrix at time $t$.}
 \label{fig:ELT}
\end{figure}

Density matrix methods are a natural choice for modeling open quantum systems and there have been a plethora of methods to capture Markovian and non-Markovian dynamics, from perturbative to numerical techniques~\cite{Breuer1999, Breuer2002, Yu2000, Shabani2005, Ishizaki2005, Chen2009, Chruscinski2010, Barchielli2012, Moix2013, Li2014, Budini2014, Montoya-Castillo2015, Breuer2016, Semina2016, Vacchini2016, deVega2017, Campbell2018, Reimer2019, Yoshioka2019, Nagy2019, Hartmann2019, Vicentini2019, HeadMarsden2019PRA, HeadMarsden2019JCP}. However, in the non-Markovian regime many of these methods struggle from the same challenges such as maintaining the positivity, and therefore physical nature, of the system density matrix. The Ensemble of Lindblad's Trajectories Method is a recently developed, formally exact method depicted in Figure~\ref{fig:ELT}, where true density matrices and auxiliary density matrices at time $t$ are represented as purple and gray respectively, and the variable $\tau$ represents the time lag~~\cite{HeadMarsden2019PRA}.  The ELT method extends the Lindbladian formalism beyond the Markovian regime through the use of an ensemble of trajectories originating from different points in the system's history~\cite{HeadMarsden2019PRA}. Due to the relationship between Lindblad operators and Kraus maps~\cite{Kraus1983}, the density matrix remains positive semidefinite for all time, and due to the ensemble average non-Markovian behavior is captured. This method was also generalized to treat systems of multiple fermions with accurate statistics~\cite{HeadMarsden2019JCP}. 
In the simplest discrete form this is mathematically equivalent to writing the density matrix as,
\begin{eqnarray}
 D(t) &=& \sum_{i = 1}^{T} \omega(\tau_i)e^{\mathcal{L}(\tau_i)}D(t-\tau_i),
 \label{main}
\end{eqnarray}
where $T$ is the maximum memory, $\omega(\tau_i)$ are the statistical weights of the $i^{\rm th}$ trajectories, and $e^{\mathcal{L}(\tau_i)}$ are the propagators. Each trajectory is a Kraus map which we can represent by the following Lindbladian trajectory,
\begin{align}
\frac{d D}{ds} &= \mathcal{L} \circ D\\
&= -i [H, D] + \sum_{i=1}^{\mathcal{M}}C_i D  C_i^{\dagger} - \frac{1}{2}\{ C_i^{\dagger} C_i,  D \},
\label{eq:lindblad}
\end{align}
where $s$ represents an effective time within the mapping, and the Lindbladian superoperator $\mathcal{L}$ can be written in terms of the system Hamiltonian $H$ and Lindbladian matrices $C_i$, which account for the interaction of the reduced density matrix with its environment through $\mathcal{M}$ different  channels~\cite{Lindblad1976}. From the properties of Kraus maps the trajectories produce positive semidefinite density matrices whose ensemble is also positive semidefinite~\cite{Kraus1983}. 

\textcolor{black}{In the continuous limit, this formulation can be related to a simplified form of the generalized master equation~\cite{HeadMarsden2019PRA}, where the traditional memory kernel is represented by the ensemble weights, the Lindbladian superoperator, and a time-derivative. In this representation, just as in traditional use of Lindblad's equation, there are multiple avenues of approach to parameter selection, including but not limited to numerical optimization, extraction from short-time propagations, and use of experimental data.}

\subsection{Quantum Algorithm}

Recent work by Hu et al. has been done to map the Lindblad equation into a dilated unitary evolution, then perform this unitary evolution on a quantum computer~\cite{Hu2020}. 
In this work, the Lindblad equation is first written in operator sum form,
\begin{equation}
    D(t) = \sum_i M_iD M_i^{\dagger},
\end{equation}
where the $M_i$ are Kraus maps corresponding to the Lindbladian channels represented by each $C_i$ in the original equation. Since Kraus maps are contraction mappings of a Hilbert space, the Sz.-Nagy dilation theorem guarantees that there exists a related unitary operator dilation in a larger Hilbert space~\cite{Hu2020}. While different orders of $d$-dilation exist and correspond to contraction mappings being applied $d$ times to a Hilbert space, the 1-dilation is sufficient to consider either populations or coherence of a two-level system. Here, we focus on the 1-dilation which produces the unitary operator,
\begin{equation}
    U_{M_i} = \begin{pmatrix}
            M_i & D_{M^{\dagger}_i}\\
            D_{M_i} & -M^{\dagger}_i
        \end{pmatrix},
\end{equation}
where $I$ is the identity matrix and $D_{M_i} = \sqrt{I - M^{\dagger}_iM_i}$. 

To perform the unitary evolution, the density matrix is reformulated in vector form and dilated to account for the added environmental degrees of freedom. The density matrix can be represented as an ensemble of vectors $\{v_j\}$ where the vectors need not be orthogonal. The $\{v_j\}$ vectors are then dilated by padding with zeroes to match the dimension of the unitary operator. In this work, only the population elements of the density matrix are considered. Given an ensemble of vectors $\{v_j\}$ and using this dilation, the population elements of the system are calculated as,
\begin{equation}
 \rho_k = \frac{1}{2}\sum_{i,j} \lvert (U_{M_i}\cdot v_j)[k]\rvert^2,
\end{equation}
where $\rho_k$ is the occupation number of the $k^{th}$ state and the summation is over the dilations of all the Kraus operators $M_i$ acting on all the vectors $v_j$. Each term can be simulated in parallel, with circuits of gate count on the order of $n^2$, where $n$ is the dimension of the system~\cite{Hu2020}.

Since the ELT method is an ensemble average of Lindbladian trajectories, the recent algorithm used to calculate Lindbladian trajectories on quantum devices can be used with the following  generalization. In previous work when practically invoking the ELT method~\cite{HeadMarsden2019PRA, HeadMarsden2019JCP}, each trajectory is written as,
\begin{equation}
    \tilde{D}(t,\tau) = e^{\mathcal{L}}D(t-\tau),
\end{equation}
where the Lindbladian term $\mathcal{L}$ is a constant in terms of time, and $D(t-\tau)$ is the density matrix initialized from $\tau$ time steps in the system's history. This equation can be directly cast into the unitary dilation framework; however, it requires storage of each $D(t-\tau)$ along the way and the production of a new circuit describing each $D(t-\tau)$, or the production of an arbitrary state at each time point. While this is possible, it would be computationally taxing and preclude the possibility for a quantum speed up. 

Alternatively, using a variable change, we can shift the time dependence from the density matrix to the Lindbladian,
\begin{equation}
    \tilde{D}(t,\tau) = e^{\mathcal{L}(t-\tau,t)}D,
\end{equation}
where the Lindbladian term $\mathcal{L}$ depends on the time lag while $D$ is a constant density matrix. Instead of requiring an arbitrary state preparation at every step, this formulation requires only variation in the \textcolor{black}{time} input to produce the Lindblad matrices and therefore the unitary dilations. The ELT increases the computational cost of a single Lindbladian trajectory in Ref. ~\citenum{Hu2020} by a multiplicative prefactor of $T$ where $T$ is the number of trajectories in the ensemble average. 

\section{Results}
\label{sec:res}

\subsection{Single Lindbladian Trajectory}

\begin{figure}[h!]
\centering
 \includegraphics[width=0.45\textwidth]{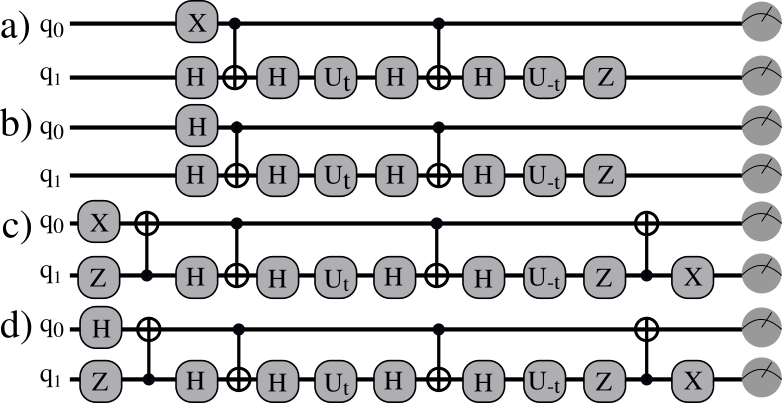}\\
 \caption{Two qubit circuits are presented for the preparation of the following states a) $U_{M_0}v_0$, b) $U_{M_0}v_1$, c) $U_{M_1}v_0$, and d) $U_{M_1}v_1$ where $X$ is the $\sigma_x$ gate, $Z$ is the $\sigma_z$ gate, $H$ is the Hadamard gate, $U_t$ is a rotation gate equivalent to $R_y(\theta)$ where $\theta = \rm{cos}^{-1}(\sqrt{e^{-\gamma t}})$, and two-qubit gates are CNOT gates. The preparations are with respect to an initial state in which both qubits are in their ground states.
 }
 \label{fig:gates}
\end{figure}

The unitary evolution matrices of a two-level system in a single amplitude damping Lindbladian channel are given by~\cite{Hu2020},
\begin{equation}
 U_{M_0} = \begin{pmatrix}
        1 & 0 & 0 & 0\\
        0 & \sqrt{e^{-\gamma t}}  & 0 & \sqrt{1-e^{-\gamma t}} \\
        0 & 0 & -1 & 0\\
        0 &  -\sqrt{1-e^{-\gamma t}} & 0 & -\sqrt{e^{-\gamma t}}
        \end{pmatrix}
\end{equation}
and
\begin{equation}
 U_{M_1} = \begin{pmatrix}
        0 & \sqrt{1-e^{-\gamma t}} & \sqrt{e^{-\gamma t}} & 0\\
        0 & 0 & 0 & 1\\
        1 & 0 & 0 & 0\\
        0 &  \sqrt{e^{-\gamma t}} & -\sqrt{1-e^{-\gamma t}} & 0
        \end{pmatrix},
\end{equation}
where $\gamma$ is the rate of decay. 
\textcolor{black}{An initial density matrix of the form,
\begin{equation}
    D(0) = \frac{1}{4}\begin{pmatrix}
            1 & 1 \\
            1 & 3\\
            \end{pmatrix},
\end{equation}
is decomposed into an ensemble of dilated vectors~\cite{Hu2020},}
\begin{equation}
v_0 = \begin{pmatrix}
    0\\
    1\\
    0\\
    0
    \end{pmatrix}
\end{equation}
and
\begin{equation}
    v_1 = \frac{1}{\sqrt{2}}\begin{pmatrix}
    1\\
    1\\
    0\\
    0
    \end{pmatrix}.
\end{equation}
\textcolor{black}{With this decomposition,} the four necessary terms for each Lindbladian trajectory are calculated through the circuits shown in Figure~\ref{fig:gates}. The circuits were constructed and verified in both Qiskit~\cite{Qiskit} and QuTip~\cite{Johansson2012, Johansson2013}. It should be noted that the two qubit states follow the conventional notation where the vector $(1, 0, 0, 0)$ represents both qubits in their ground states, $(0, 1, 0, 0)$ represents $q_0$ in its excited state and $q_1$ in its ground state, $(0, 0, 1, 0)$ represents $q_0$ in its ground state and $q_1$ in its excited state, and $(0, 0, 0, 1)$ represents both qubits in their excited states.

\begin{figure}[h!]
\centering
 \includegraphics[width=0.5\textwidth]{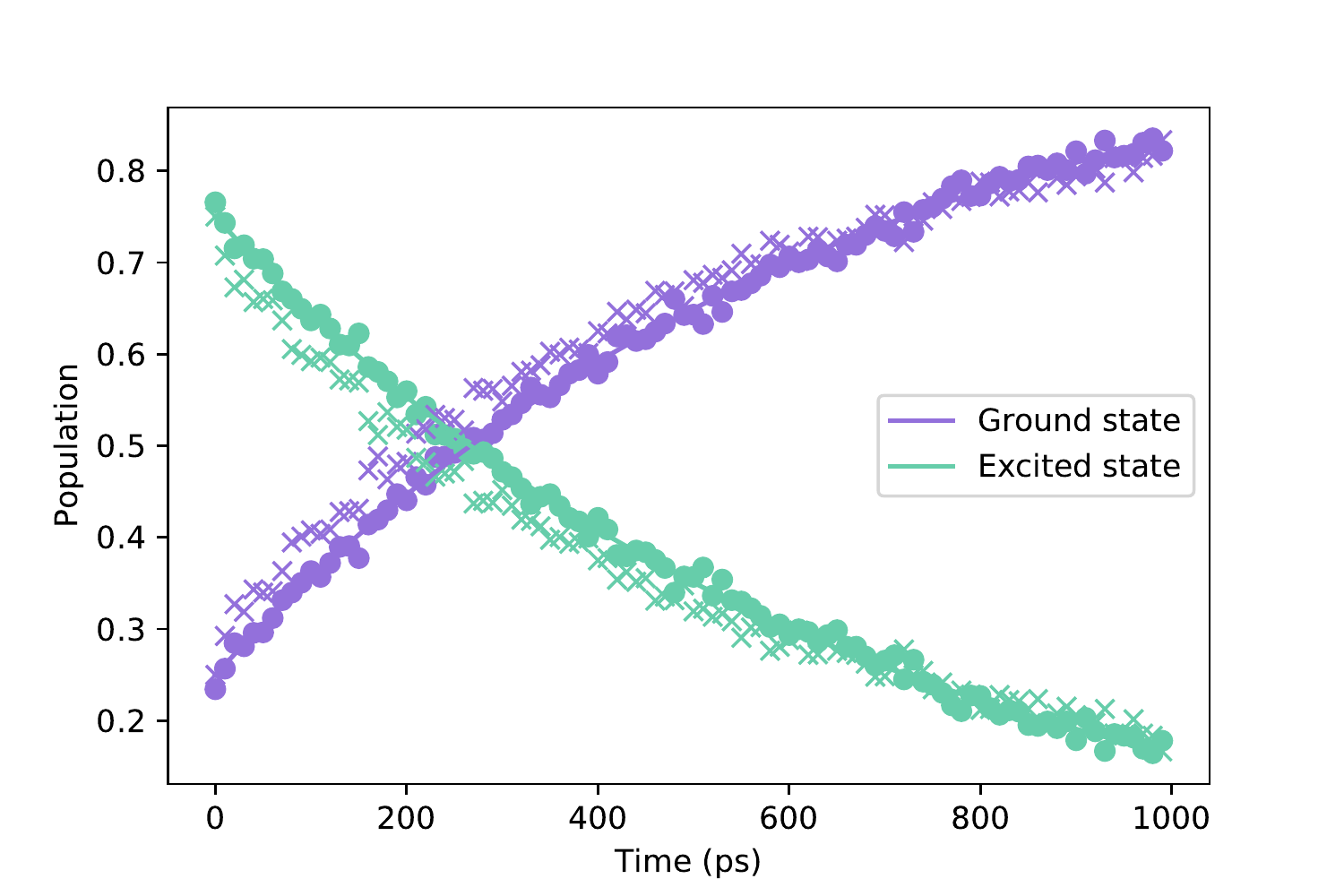}\\
 \caption{Two level system in an amplitude damping channel where purple represents the ground state population and teal the excited state. The solid lines are the solution on a classical device, the dots are generated using the IBM Qiskit simulator,~\cite{Qiskit} and the x's are generated using IBM's London device~\cite{IBM}.}
 \label{fig:lindblad}
\end{figure}

Using these circuits, we consider a two-level system in an amplitude damping channel with a system decay rate of  $\gamma = 1.52\cdot 10^9s^{-1}$. Using the solution from the Lindblad equation on a classical device, IBM's Qiskit simulator~\cite{Qiskit}, and IBM's London device~\cite{IBM} the ground and excited state populations, in purple and teal respectively, are shown in Figure~\ref{fig:lindblad}. The solid lines represent the solution to the Lindblad equation on a classical device, the dots represent the result from the simulator, and the x's result from the \textcolor{black}{5-qubit} London device. \textcolor{black}{Both the simulator and device data agree well with the classical solution, producing accurate dynamics for a two-level system under Markovian conditions.}

\subsection{Ensemble of Lindbladian Trajectories}

Having verified this \textcolor{black}{algorithm for a single trajectory of the} Lindblad equation, we extend these circuits to treat the damped Jaynes-Cummings model, which consists of a single excitation in a two-level system coupled to a reservoir of harmonic oscillators~\cite{Jaynes1963, Stenholm2013,Breuer2002, Breuer1999}. \textcolor{black}{ This system is both exactly solvable and known to demonstrate non-Markovian behavior in the strong coupling and detuned regimes, making it an excellent benchmarking system for open quantum system methods\cite{Jaynes1963, Breuer2002, HeadMarsden2019PRA}.} The Hamiltonian is given by,
\begin{equation}
    \hat{H}_{JC} = \hbar \omega \hat{a}^\dagger\hat{a} + \frac{1}{2}\hbar \omega_0\hat{\sigma}_z + \hbar\lambda(\hat{\sigma}_+\hat{a} + \hat{\sigma}_-\hat{a}^{\dagger}),
\end{equation}
where $\omega_0$ is the system's transition frequency, $\lambda$ is inversely proportional to the reservoir correlation time, $\hat{a}^{\dagger}$ and $\hat{a}$ are the creation and annihilation operators, and $\hat{\sigma}_{x,y,z}$ are the Pauli spin operators with $\hat{\sigma}_{\pm} = \hat{\sigma}_x \pm \hat{\sigma}_y$. In lieu of the cavity mode treatment, the bath spectral density is used~\cite{Breuer2002},
\begin{equation}
    J(\omega) = \frac{1}{2\pi}\frac{\gamma\lambda^2}{(\omega_0-\Delta-\omega)^2 + \lambda^2},
\end{equation}
where $\omega_0$ is the system transition frequency, $\omega$ the bath frequency, $\Delta$ the detuning, and $\lambda$ is related to the system-bath coupling strength.

\begin{figure}[h!]
\centering
 \includegraphics[width=0.5\textwidth]{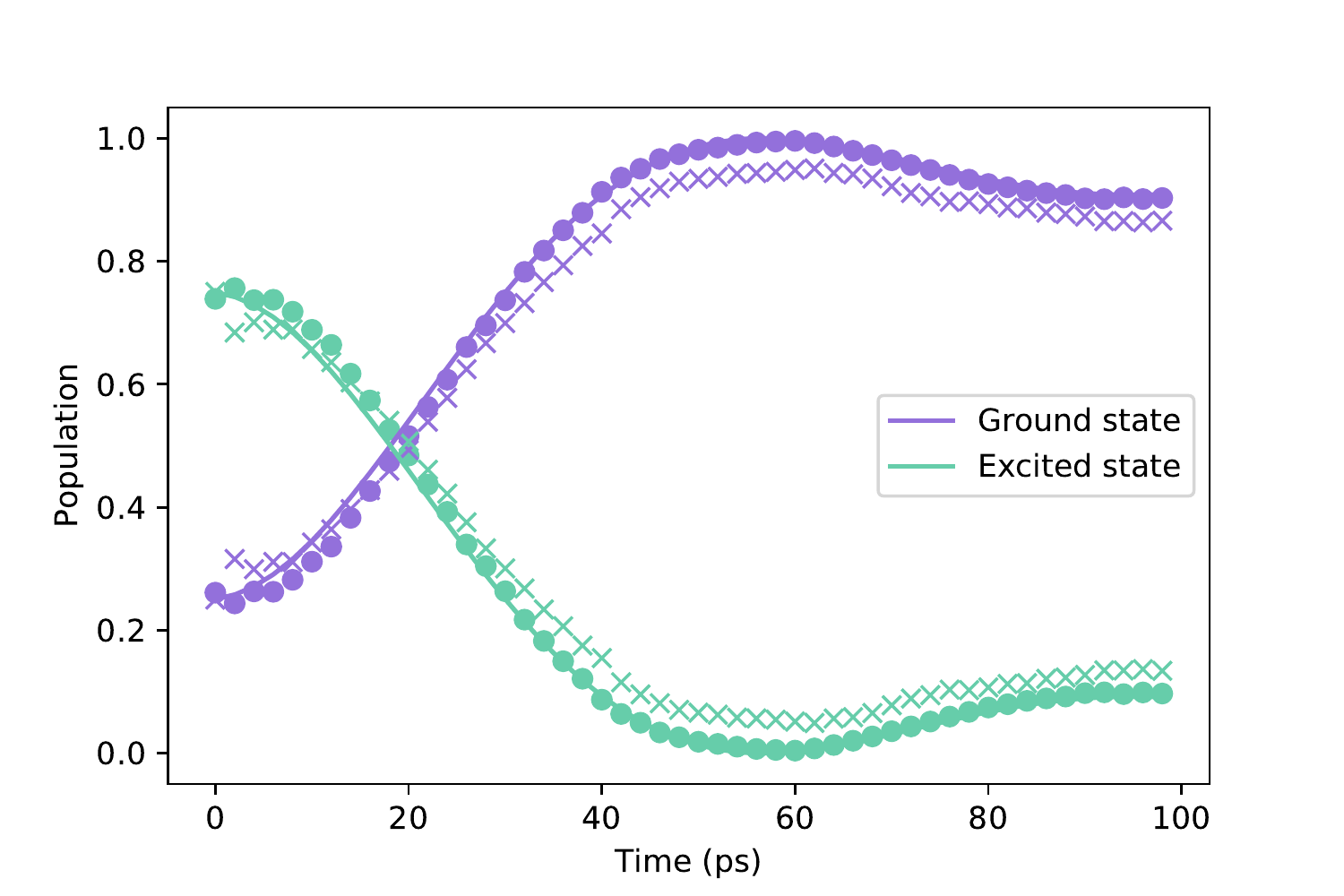}\\
 \caption{Populations of the Jaynes-Cummings model in the strong coupling regime, $\lambda = 0.2\gamma$, where purple represents the ground state population and teal the excited state. The solid lines are the exact solution, the dots are generated using the IBM Qiskit simulator~\cite{Qiskit}, and the x's are generated using IBM's London device~\cite{IBM}.}
 \label{fig:JC_strong}
\end{figure}

First we consider the strong coupling case where the bath relaxation parameter is given by $\lambda = 0.2\gamma$. In this regime, a single Lindbladian trajectory fails to capture accurate dynamics and a more involved method is required~\cite{Breuer1999, Breuer2002}. The populations of the Jaynes-Cummings model in the strong coupling regime are shown in Figure~\ref{fig:JC_strong} using the ELT-QC method with weights numerical optimized as compared to the exact solution using Maple~\cite{Maple}, as was done in Ref.~\citenum{HeadMarsden2019PRA}. The solid lines represent the exact solution while the dots and x's are results from quantum simulation using the circuits shown in Fig.~\ref{fig:gates} using IBM's simulator and London device respectively~\cite{Qiskit, IBM}.

The ELT on a quantum simulator and device agrees well with the exact solution, demonstrating ability of the ELT-QC algorithm to accurately capture dynamics in the strong coupling regime. 

\begin{figure}[h!]
\centering
 \includegraphics[width=0.5\textwidth]{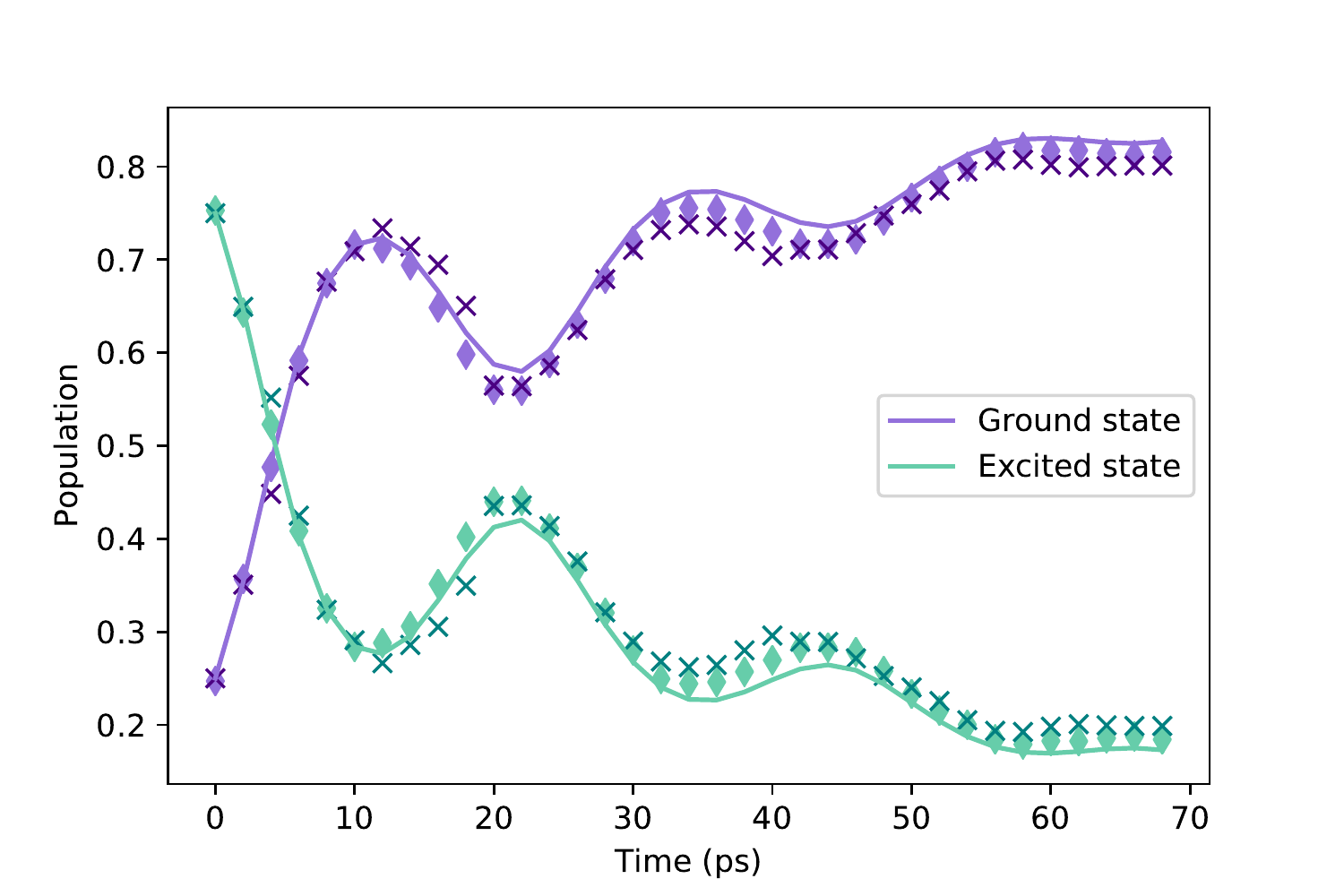}\\
 \caption{Populations of the Jaynes-Cummings model in the strong coupling and detuned regime, $\lambda = 0.3\gamma$ and $\Delta = 2.4\gamma$, where purple represents the ground state population and teal the excited state. The solid lines are the exact solution, the dots are generated using the IBM Qiskit simulator~\cite{Qiskit}, and the x's are generated using IBM's London device~\cite{IBM}.}
 \label{fig:JC_detuned}
\end{figure}

\textcolor{black}{Next we consider} the strong coupling and detuned case, where $\Delta = \omega - \omega_0 \neq 0$. \textcolor{black}{In this regime, }the bath relaxation parameter is given by $\lambda = 0.3\gamma$ and the detuning by $\Delta = 2.4\gamma$. The \textcolor{black}{time-evolution of the }populations are shown in Figure~\ref{fig:JC_detuned} where lines represent the exact result and the dots and x's represent results from quantum simulation using the circuits shown in Fig.~\ref{fig:gates} using IBM's simulator and London device respectively.
\textcolor{black}{These results are in excellent agreement with the exact solution, demonstrating that the} ELT-QC algorithm captures accurate dynamics in the detuned regime for the Jaynes-Cummings model.

\section{Conclusions and Outlook} 
\label{sec:conc}

Few algorithms exist for the treatment of open quantum systems on quantum devices, and existing algorithms are either restricted to Markovian systems or inconsistent with complete positivity. Our work presents an algorithm inspired by the Ensemble of Lindblad's Trajectories method~\cite{HeadMarsden2019PRA} using an efficient dilation theorem~\cite{Hu2020} which allows for the treatment of non-Markovian dynamics of open quantum systems on a quantum device. The ELT-QC method is benchmarked on the Jaynes-Cummings model in the strong coupling and detuned cases on \textcolor{black}{both a quantum simulator and real 5-qubit quantum computer}, showing excellent agreement with the exact dynamics. \textcolor{black}{While the Jaynes-Cummings model is a benchmark system, these results show our algorithm's ability to capture a wide variety of open quantum system dynamics accurately with tractable scaling. These regimes include strictly Markovian, or dissipative, dynamics and non-Markovian dynamics induced by either strong or detuned system-environment coupling.} 

Moreover, this quantum algorithm retains the significant advantages of its classical counterpart including an exact treatment of non-Markovian dynamics and complete positivity of the density matrices, meaning that the density matrices remain positive semidefinite with non-negative probabilities for all time. The ELT-QC algorithm offers a path towards polynomial time-evolution of an electronic system in the presence of a complex environment. An accurate yet tractable description of open quantum systems on quantum devices has a myriad of significant applications from catalytic chemistry and correlated materials physics to descriptions of hybrid quantum systems and spin systems. \\

\begin{acknowledgments}
\noindent \textbf{Acknowledgments} 
The authors thank Dr. Evelyn M. Goldfield for the constructive feedback on the manuscript. This work is partially supported by the U.S. Department of Energy, Office of Science, Basic Energy Sciences (BES), Materials Sciences and Engineering Division under FWP ERKCK47 `Understanding and Controlling Entangled and Correlated Quantum States in Confined Solid-state Systems Created via Atomic Scale Manipulation'. KHM, SK and PN acknowledge support from the Harvard Physical Sciences Accelerator Award and Harvard Quantum Initiative Seed Grant. DAM gratefully acknowledges the U.S. Department of Energy, Office of Science, Basic Energy Sciences (BES) under Grant No. DE-SC0019215,  the National Science Foundation (NSF) under Grant No. CHE-1152425, and  the U.S. Army Research Office Grant No. W911NF-16-1-015.  P.N. is a Moore Inventor Fellow supported through Grant No. GBMF8048 from the Gordon and Betty Moore Foundation. We acknowledge the use of IBM Quantum services for this work. The views expressed are those of the authors, and do not reflect the official policy or position of IBM or the IBM Quantum team.

\end{acknowledgments}

\bibliography{main}

\end{document}